\documentclass[sigconf]{acmart}
\usepackage{bm}
\usepackage{array}
\usepackage{subfigure}

\usepackage{multirow}

\AtBeginDocument{%
  }

\copyrightyear{2023}
\acmYear{2023}
\setcopyright{rightsretained}
\acmConference[e-Energy '23]{The 14th ACM International Conference on Future Energy Systems}{June 20--23, 2023}{Orlando, FL, USA}
\acmBooktitle{The 14th ACM International Conference on Future Energy Systems (e-Energy '23), June 20--23, 2023, Orlando, FL, USA}
\acmDOI{10.1145/3575813.3595199}
\acmISBN{979-8-4007-0032-3/23/06}

\begin{document}

\title{A Scalable Bilevel Framework for Renewable Energy Scheduling }

\author{Dongwei Zhao}
\email{zhaodw@mit.edu}
\affiliation{%
  \institution{Massachusetts Institute of Technology}
  \streetaddress{P.O. Box 1212}
  \city{Cambridge}
  \state{MA}
  \country{USA}
  \postcode{43017-6221}
}

\author{Vladimir Dvorkin}
\email{dvorkin@mit.edu}
\affiliation{%
  \institution{Massachusetts Institute of Technology}
  \streetaddress{P.O. Box 1212}
  \city{Cambridge}
  \state{MA}
  \country{USA}
  \postcode{43017-6221}
}

\author{Stefanos Delikaraoglou}
\email{sdelikaraoglou@gmail.com}
\affiliation{%
  \institution{Massachusetts Institute of Technology}
  \city{Cambridge}
  \state{MA}
  \country{USA}
  \postcode{43017-6221}
}

\author{Alberto J. Lamadrid L.}
\email{ajlamadrid@lehigh.edu}
\affiliation{%
  \institution{Lehigh University}
  \city{Bethlehem}
  \state{PA}
  \country{USA}
}

\author{Audun Botterud}
\email{audunb@mit.edu}
\affiliation{%
  \institution{Massachusetts Institute of Technology}
  \city{Cambridge}
  \state{MA}
  \country{USA}}

\begin{abstract}
 Accommodating the uncertain and variable renewable energy sources (VRES) in electricity markets requires sophisticated and scalable tools to achieve market efficiency.  To account for the uncertain imbalance costs in the real-time market while remaining compatible with the existing sequential market-clearing structure, our work adopts an uncertainty-informed adjustment toward the VRES contract quantity scheduled in the day-ahead market. This mechanism requires solving a bilevel problem, which is computationally challenging for practical large-scale systems. To improve the scalability, we propose a technique based on  strong duality and McCormick envelopes, which relaxes the original problem to linear programming. We conduct numerical studies on both IEEE 118-bus and 1814-bus NYISO systems. Results show that the proposed relaxation can achieve good performance in accuracy  (0.7\%-gap in the system cost wrt. the least-cost stochastic clearing benchmark) and scalability (solving the NYISO system in minutes). Furthermore, the benefit of this bilevel VRES-quantity adjustment is more significant under higher penetration levels of VRES (e.g., 70\%), under which the system cost can be reduced substantially compared to a myopic day-ahead offer strategy of VRES.
\end{abstract}



\maketitle

\section{Introduction}

The appeal for a transition to zero-carbon power systems has motivated the large-scale deployment of variable renewable energy sources (VRES). From 2010 to 2020, the globally installed solar PV capacity increased from about 40 GW to 760 GW, while the wind capacity grew from about 200 GW to 740 GWh \cite{renreport2021}. In many regions, this transition takes place in restructured electricity markets, which were initially designed according to the technical and economic properties of dispatchable generators, usually fossil-fueled. As a result, the ability of the current market design to efficiently accommodate large shares of VRES is called into question.

Current short-term electricity markets are typically organized into two main settlements \cite{fundamentals}: a day-ahead (DA) market  cleared before actual operations to establish initial generation schedules and prices, and a real-time (RT) market that runs close to actual delivery in order to compensate for any imbalances from the DA schedule. In this sequential settlement process, the uncertainty information about VRES generation in the DA market is usually summarized in a single-valued point forecast \cite{nyisowindrule},  typically the conditional expectation of the predictive probability distribution. This myopic view of forecast uncertainty leads to imperfect coordination between the DA and RT operations, since the DA schedule does not account for the re-dispatch cost under generation uncertainty.

Aiming to improve scheduling efficiency accounting for system uncertainties,  the works \cite{morales2012pricing, Exizidis2019market, Kazempour2018Market} proposed alternative dispatch approaches which co-optimized DA and RT operations to minimize total expected system costs. Although these models attain perfect temporal coordination between the DA and RT stages, they are not readily compatible with the existing market structure in which DA and RT markets are cleared sequentially and separately. Morales \textit{et al.} \cite{morales2014electricity} proposed an adjustment of day-ahead VRES quantities based on bilevel models. Such a framework can account for the imbalance costs in the RT market while maintaining the sequential market-clearing structure.  Several follow-up works applied this bilevel approach to other reserve and energy problems\cite{viafora2020dynamic,delikaraoglou2019optimal,dvorkin2018setting}. However, these works solved the bilevel problem by formulating a single mixed-integer linear programming (MILP) problem, which was only tested on small-scale systems, e.g., a 24-bus system \cite{morales2014electricity}. The MILP problems are hard to solve for  large-scale systems, e.g., NYISO, and thus may fail to support practical market operations.

In contrast to the literature above, following the bilevel optimization model,  our work aims to \textit{efficiently compute the optimal quantity adjustment for VRES in the DA market for a practical large-scale system.}  We summarize the key contributions  as follows.

 \textit{Efficient algorithm for large-scale systems:} To solve the bilevel problem of VRES quantity adjustment,  the conventional method is to reformulate a MILP problem using KKT conditions. We find that this approach cannot solve the 1814-bus NYISO system within two hours. To resolve this scalability issue,  we propose a technique based on the strong duality of linear programming and McCormick-envelope relaxation. This method is computationally efficient as it only requires solving a linear program, which can produce solutions for the NYISO system in minutes.

 \textit{Performance and benefit:} We conduct numerical studies on IEEE 118-bus and 1814-bus NYISO systems. The results show that the proposed relaxation technique can achieve accuracy (0.7\%-gap in the system cost wrt. the least-cost stochastic dispatch benchmark) and scalability (solving the NYISO system in minutes). Besides, the benefit of the bilevel VRES-quantity adjustment is more significant under high levels of VRES (e.g., 70\%), where the system cost can be reduced by over 15\%  compared to the myopic strategy.   


\section{Two-settlement market clearing}\label{sec:market}
This section provides optimization models for conventional DA and RT market dispatch.  In terms of modeling assumptions, the network topology is included considering linear DC power flows. Energy supply functions are linear, and all generators behave as price takers. For simplicity, the sole source of uncertainty is VRES power production, and system demand is inelastic with a high value of lost load (VoLL). We focus on hourly economic dispatch, which will be generalized to the unit commitment problem in future work. We define all the notation in Appendix A.

\subsection{Day-ahead market} 
The DA market clearing problem minimizes the total generation costs $f^{\text{DA}}$ of all the conventional units. 
\begin{subequations} \label{eq:da}
\allowdisplaybreaks
	\begin{align}
		\min ~&f^{\text{DA}}(\Phi^{\text{DA}}):= \sum_{i \in \mathcal{I}} C_{i}  P_{i}^{\text{C}}  \\
		\text{s.t.} 
		~&\sum_{i \in \mathcal{I}_{n}}\!\!P_{i}^{\text{C}} \!+ \!\!\!\!\sum_{k \in \mathcal{K}_{n}}\!\!P_{k}^{\text{W}} - L_{n} -\!\!\!\!\sum_{m:(n,m)\in\Lambda}\!\!\!\!\!\!\!\!\frac{\delta_{n}^{\text{DA}}-\delta_{m}^{\text{DA}}}{x_{nm}} = 0, 
\forall n \in \mathcal{N},    ~~~~[\lambda_n^b]\label{eq:dabalance}\\
		&-\overline{F}_{nm}\leq \frac{\delta_{n}^{\text{DA}}-\delta_{m}^{\text{DA}}}{x_{nm}} \leq \overline{F}_{nm}, ~ \forall (n,m) \in \Lambda,[\underline{\lambda}_{nm},\overline{\lambda}_{nm}]\label{eq:daline}\\		&\underline{P}_{i}^{\text{C}} \leq P_{i}^{\text{C}} \leq \overline{P}_{i}^{\text{C}}, ~\forall i\in \mathcal{I}, \hspace{13ex} [\underline{\lambda}_i^\text{C}, \overline{\lambda}_i^\text{C}]\label{eq:dac}\\
    		&0 \leq P_{k}^{\text{W}} \leq {{W}_{k}}, ~ \forall k\in \mathcal{K},\hspace{12.5ex}[\underline{\lambda}_k^\text{W}, \overline{\lambda}_k^\text{W}]\label{eq:daw} \\
  \text{var}:~&\Phi^{\text{DA}} = \{P_{i}^{\text{C}}, \forall i; P_{k}^{\text{W}}, \forall k; \delta_{n}^{\text{DA}}, \forall n \}.\notag
	\end{align}
\end{subequations}
The decision variable set $\Phi^{\text{DA}}$ comprises the DA schedule  for each conventional and VRES unit as well as voltage angles at each bus. Constraint \eqref{eq:dabalance} ensures power balance. Constraints \eqref{eq:daline}-\eqref{eq:daw} enforce the  power flow and generators' schedule within the capacity limits. In constraint \eqref{eq:daw}, $\bm{W}=(W_k,\forall k\in \mathcal{K})$ is the VRES quantity offer, which can be empirically set at the mean value of the forecast or optimized in the later bilevel model. We denote by $\mathcal{X}^{\text{DA}}(\bm{W})$ the constraint set constructed by  \eqref{eq:dabalance}-\eqref{eq:daw} under the VRES offer $\bm{W}$. We list the dual variables $\bm{\lambda}$ corresponding to each constraint and denote the optimal DA schedule as $\Phi^{\text{DA}*}$.

\subsection{Real-time market}

Getting closer to RT operation, any deviation from the  DA schedule $\Phi^{\text{DA}}$ has to be covered by balancing actions. For a realization $\omega\in \Omega$ of random production {$\widetilde{W}_{k\omega}$}, the system operator decides the optimal re-dispatch by minimizing the redispatch cost $f_\omega^{\text{RT}}$ as follows.
\begin{subequations} 
\allowdisplaybreaks
	\begin{align}
\min~ & f_\omega^{\text{RT}}(\Phi_\omega^{\text{RT}}):=\sum_{i \in \mathcal{I}} \! C_{i}^U r_{i\omega}^{\text{U}}\!- \!C_{i}^D r_{i\omega}^{\text{D}} \!+\!\!\!\sum_{n \in \mathcal{N}}\! C^{\text{VoLL}} L_{n\omega}^{\text{sh}} \\
		\text{s.t.}~&  \sum_{i \in \mathcal{I}_{n}} \left( r_{i\omega}^{\text{U}}-r_{i\omega}^{\text{D}} \right)
		\!+\! \sum_{k \in \mathcal{K}_{n}}\!\left(\widetilde{W}_{k\omega} - P_{k}^{\text{W}} - P_{k\omega}^{\text{W,cr}} \right)   + L_{n\omega}^{\text{sh}}\notag \\
		&- \hspace{-2ex}
		\sum_{m:(n,m)\in\Lambda} \hspace{-3ex}\frac{\delta_{n\omega}^{\text{RT}}-\delta_{n}^{\text{DA}}-\delta_{m\omega}^{\text{RT}}+\delta_{m}^{\text{DA}}}{x_{nm}} = 0,  \forall n \in \mathcal{N}, \label{eq:rtbalance}\\
		&0\leq r_{i\omega}^{\text{U}} \leq \overline{P}_{i}-P_{i}^{\text{C}}, ~\forall i \in \mathcal{I}, \label{eq:rtup}\\
& 0\leq r_{i\omega}^{\text{D}} \leq P_{i}^{\text{C}}-\underline{P}_{i}^{\text{C}},~ \forall i \in \mathcal{I},\label{eq:rtdown}\\
		& -\overline{F}_{nm}\leq \frac{\delta_{n\omega}^{\text{RT}}-\delta_{m\omega}^{\text{RT}}}{x_{nm}} \leq \overline{F}_{nm}, ~\forall (n,m) \in \Lambda, \label{eq:rtline} \\
		&0 \leq P_{k\omega}^{\text{W,cr}} \leq \widetilde{W}_{k\omega}, ~\forall k \in \mathcal{K},\label{eq:rtw}  \\
		&0 \leq L_{n\omega}^{\text{sh}} \leq L_{n}, ~ \forall n \in \mathcal{N},\label{eq:rtlost} \\
  \text{var}:~&\Phi_\omega ^{\text{RT}} = \{r_{iw}^{\text{U}},r_{iw}^{\text{D}}, \forall i; P_{k\omega}^{\text{W,cr}}, \forall k; \delta_{n\omega}^{\text{RT}},L_{n\omega}^{\text{sh}},\forall n \}.\notag
	\end{align}
\end{subequations}
The decision variable set $\Phi_\omega ^{\text{RT}}$ comprises the redispatch variables as well as real-time voltage angles at each bus. Constraint \eqref{eq:rtbalance} ensures power re-balance. Constraints \eqref{eq:rtup}-\eqref{eq:rtlost} enforce the upward and downward adjustments, transmission power flow, VRES curtailment, and shed load within the capacity limits.  We denote by $\mathcal{X}^{\text{RT}}(\Phi ^{\text{DA}})$ the constraint set constructed by  \eqref{eq:rtbalance}-\eqref{eq:rtlost}, which is coupled with DA schedule $\Phi ^{\text{DA}}$. Note that both optimization problems for the DA and RT dispatch are linear programming (LP) problems.

\section{Bilevel quantity adjustments}\label{section:bilevel}
For sequential DA and RT markets, we develop an uncertainty-informed bilevel optimization model to find the cost-optimal quantity offers of VRES, which improves the coordination between the two markets in terms of expected system costs. We also introduce two dispatch benchmarks for the proposed model. 

\subsection{Bilevel optimization model}
We introduce a bilevel model \textit{BiD}. In the upper level, the system operator will decide the day-ahead bidding quantity $W_k$ for each VRES producer $k$. In the lower level, the DA schedule $\Phi^{\text{DA}}$ will be optimized given the bidding quantity $\bm{W}$ announced from the upper problem. Constrained by the DA schedule $\Phi^{\text{DA}}$ in the lower level, the system operator will also decide the real-time redispatch $\Phi^{\text{RT}}$ in addition to the day-ahead VRES bidding quantity $\bm{W}$ in the upper level. The upper level aims to minimize the system cost including the DA cost and RT expected costs. We formulate such a bilevel optimization problem \textit{BiD} as follows.

\setlength{\fboxsep}{6pt}
\setlength{\fboxrule}{1pt}
\fcolorbox{black}{gray!2}{%
   \parbox{0.88\columnwidth}{%
  
\textbf{Problem \textit{BiD}: Bilevel optimization problem for day-ahead VRES quantity offer}
\begin{subequations} \label{prob:bilevel_clearing}	
	\begin{align*}
	\hspace{-3ex}	\underset{\Phi^{\text{RT}}\bigcup \bm{W}}{\text{min}} ~&
		f^{\text{DA}}(\Phi^{\text{DA}*}) + \mathbb{E}_{\omega\in \Omega}\left[f_\omega^{\text{RT}}(\Phi_\omega^{\text{RT}})\right] \\	
	\text{s.t.} ~& \Phi_\omega^{\text{RT}}\in \mathcal{X}_\omega^{\text{RT}}(\Phi^{\text{DA}*}),~\forall \omega \in \Omega,\\
		&\Phi^{\text{DA}*}  \in \text{arg}
		\left\{\!\begin{aligned}
			\underset{\Phi^{\text{DA}}}{\text{min}} ~&f^{\text{DA}}(\Phi^{\text{DA}})\\
			\text{s.t.} ~& \Phi^{\text{DA}}\in \mathcal{X}^{\text{DA}}(\bm{W})\\
		\end{aligned}\right\}. 
	\end{align*}
 \vspace{-2ex}
\end{subequations}
      }  
    }
    
\vspace{0.6ex}
After obtaining the optimal VRES offer quantity $\bm{W}^*$, the system operator will first clear the DA market without any information  on the real-time stage, and then sequentially clear the RT market, as in the actual practice of system operators.

We consider two dispatch  benchmarks for \textit{BiD}. (i) \textit{Myopic dispatch (MyD):} Each VRES producer $k$ offers the bidding quantity at the expected value of the forecast, i.e.,   $W_k=\mathbb{E}_{\omega \in \Omega}[\widetilde{W}_{k\omega}]$. Then, the DA and RT markets are cleared sequentially. (ii) \textit{Stochastic dispatch (StD):}
The system operator co-optimizes the day-ahead schedule and real-time re-dispatch by minimizing the total expected costs, which we explain in Appendix \ref{sec:benchmark}. We denote the optimal system cost under  \textit{StD} as $S^{\text{StD}}$, and  the system costs  under \textit{BiD} and \textit{MyD} as  $S^{\text{BiD}}$ and $S^{\text{MyD}}$, respectively. The system costs always satisfy  $S^{\text{MyD}}\geq S^{\text{BiD}}\geq S^{\text{StD}}$.
 The two benchmarks serve as the upper bound and lower bound for the proposed bilevel model. We show more details of benchmarks in Appendix \ref{sec:benchmark}.

\section{Solution method}\label{section:solution}

The bilevel problem \textit{BiD} is non-convex and challenging to solve. The conventional method is to replace the lower-level problem using KKT conditions and transform the problem into an MILP problem \cite{morales2014electricity}. This method can work well on a small-scale system (e.g., IEEE 118-bus system), but it cannot solve a large-scale system (e.g., NYISO) in a reasonable time. To resolve this scalability issue,  we propose a method based on the strong duality of LP and  McCormick-envelope relaxation.

\subsection{Strong duality transformation}
Based on the strong duality of LP, we establish a set of new constraints equivalent to KKT conditions.

For the low-level LP problem, the KKT conditions include primary feasibility, dual feasibility, stationary conditions, and complementarity constraints. For LP, the strong-duality condition is equivalent to complementarity constraints \cite{boyd2004convex}. We list these constraints in the following: (i) Primary feasibility: \eqref{eq:dabalance}-\eqref{eq:daw}; (ii) Dual feasibility: $\tilde{\bm{\lambda}}\geq0$, where $\tilde{\bm{\lambda}}$ includes the set of dual variables $\bm{\lambda}$ except those associated with  \eqref{eq:dabalance}.
(iii) Stationary conditions:
        \begin{subequations}
        \allowdisplaybreaks
    \begin{align}
        &C_i- \lambda_n^b-\underline{\lambda}_i^C+\overline{\lambda}_i^C=0,~\forall n \in \mathcal{N},\forall i \in \mathcal{I}_n,\label{eq:stata}\\
    &- \lambda_n^b-\underline{\lambda}_k^W+\overline{\lambda}_k^W=0,~\forall n \in \mathcal{N},\forall k \in \mathcal{K}_n,\\
        &\hspace{-2ex}\sum_{m:(n,m)\in\Lambda}\hspace{-3ex}\frac{-\lambda_n-\underline{\lambda}_{nm}+\overline{\lambda}_{nm}}{x_{nm}}=\hspace{-3ex}\sum_{m:(m,n)\in\Lambda}\hspace{-2ex}\frac{-\lambda_m-\underline{\lambda}_{mn}+\overline{\lambda}_{mn}}{x_{mn}},\forall n \in \mathcal{N}.\label{eq:statz}
    \end{align}
    \end{subequations}
(iv) Strong-duality condition:
    \begin{align}
        f^{\text{DA}}(\Phi^{\text{DA}}) =  g^{\text{DA}}(\bm{\lambda}), \label{eq:strongdual}
    \end{align}
    where the dual objective function is defined as 
        \begin{align}
     g^{\text{DA}}(\bm{\lambda}):=&\sum_{n\in \mathcal{N}} \lambda_n^b\cdot L_n- \sum_{n,m\in \mathcal{N}} (\underline{\lambda}_{nm}+\overline{\lambda}_{nm})\cdot \overline{F}_{nm}\notag\\
     &\hspace{-5ex}-\sum_{i\in \mathcal{I}} \overline{\lambda}_i^C \cdot \overline{P}_i^C+\sum_{i\in \mathcal{I}} \underline{\lambda}_i^C \cdot \underline{P}_i^C-\sum_{k\in \mathcal{K}} \overline{\lambda}_k^W \cdot W_k.\label{eq:dualobj}
    \end{align}


However, the function $g^{\text{DA}}(\bm{\lambda})$ has the bilinear term $\overline{\lambda}_k^W \cdot W_k$. Next, we adopt the McCormick envelope to relax this bilinear item.

\subsection{McCormick-envelope relaxation}
We will formulate the McCormick-envelope relaxation. We  show how we choose the bounds for the envelope in Appendix \ref{sec:bound}.

 We let $z_k=\overline{\lambda}_k^W \cdot W_k$ in \eqref{eq:dualobj} and change $g^{\text{DA}}(\bm{\lambda})$ into $g^{\text{DA}}(\bm{\lambda},\bm{z})$, which transforms \eqref{eq:strongdual}  into   
\vspace{-0.5ex}
\begin{align}
        f^{\text{DA}}(\Phi^{\text{DA}}) =  g^{\text{DA}}(\bm{\lambda},\bm{z}). \label{eq:strongdualn}
    \end{align}

If we have the bounds $\alpha_k^\lambda\leq \overline{\lambda}_k^W \leq \beta_k^\lambda$ and $\alpha_k^W\leq  W_k \leq \beta_k^W$, the McCormick envelope gives the convex relaxation for  $z_k$ \cite{mccormick1976computability}: 
\vspace{-1ex}
\begin{subequations}
\allowdisplaybreaks
\begin{align}
&z_k \geq \alpha_k^\lambda \cdot W_k+ \alpha_k^W \cdot \overline{\lambda}_k^W  -\alpha_k^\lambda \alpha_k^W,~\forall k \in \mathcal{K},\label{eq:mca}\\
&z_k \geq \beta_k^\lambda \cdot W_k+ \beta_k^W\cdot \overline{\lambda}_k^W  -\beta_k^\lambda \beta_k^W,~\forall k \in \mathcal{K},\\
&z_k \leq \beta_k^\lambda \cdot W_k+ \alpha_k^W\cdot \overline{\lambda}_k^W  -\beta_k^\lambda \alpha_k^W,~\forall k \in \mathcal{K},\\
&z_k \leq \alpha_k^\lambda \cdot W_k+\beta_k^W\cdot \overline{\lambda}_k^W  -\alpha_k^\lambda \beta_k^W,~\forall k \in \mathcal{K}. \label{eq:mcz}
\end{align}
\end{subequations}
This leads to the following relaxed  problem  for Problem \textit{BiD}.

\setlength{\fboxsep}{6pt}
\setlength{\fboxrule}{1pt}
\fcolorbox{black}{gray!2}{%
   \parbox{0.88\columnwidth}{%
\textbf{
\hspace{-1ex}Problem \textit{BiD-McCormick}: Relaxed problem}
\vspace{-0.8ex}
\begin{subequations}	
	\begin{align}
		\min  ~&
		f^{\text{DA}}(\Phi^{\text{DA}}) + \mathbb{E}_{\omega\in \Omega}\left[f_\omega^{\text{RT}}(\Phi_\omega^{\text{RT}})\right] \notag \\	
	\text{s.t.} ~& \Phi_\omega^{\text{RT}}\in \mathcal{X}_\omega^{\text{RT}}(\Phi^{\text{DA}}),~\forall \omega \in \Omega,\notag\\
		&\Phi^{\text{DA}}\in \mathcal{X}^{\text{DA}}(\bm{W}),\notag\\
  &\tilde{\bm{\lambda}}\geq 0, ~\eqref{eq:stata}-\eqref{eq:statz}, ~\eqref{eq:strongdualn},~\eqref{eq:mca}-\eqref{eq:mcz}\notag\\
  \text{var}:~&\Phi^{\text{DA}},\Phi^{\text{RT}}, \bm{W}, \bm{\lambda},\bm{z}. \notag
	\end{align}
\end{subequations}
\vspace{-3.5ex}
}}
\vspace{1ex}

The relaxed problem \textit{BiD-McCormick} is a LP problem that can be efficiently solved even on large-scale systems. Recall that after obtaining the quantity solution $\bm{W}^*$ from \textit{BiD-McCormick}, the system operator will first clear the DA market given $\bm{W}^*$, and then sequentially clear the RT market, based on which we calculate the system cost under \textit{BiD-McCormick}.

\section{Numerical studies}\label{section:simulation}

We use the IEEE 118-bus system \cite{ieee118} to demonstrate the accuracy of the proposed \textit{BiD-McCormick} formulation and adopt the NYISO system \cite{greene2022} to show scalability. We simulate on a MacBook Pro (2020) with a 2.3 GHz Quad-Core Intel Core i7 processor. We use the solver Mosek to solve LP and Gurobi to solve MILP  in Julia/JuMP. 

\subsection{Case of the IEEE 118-bus system}

For the 118-bus system, we consider 14 spatially distributed wind farms. We randomly generate 20 forecast scenarios of wind energy following the truncated normal distribution within the wind capacity.  We demonstrate that \textit{BiD-McCormick} can achieve good accuracy and tightness. Also, the benefit of \textit{BiD} is more significant under higher VRES integration.

\subsubsection{Good performance of \textit{BiD-McCormick}}

\begin{figure}[t]
	\centering
	\hspace{-2ex}
	\subfigure[]{
		\raisebox{-2mm}{\includegraphics[width=1.67in]{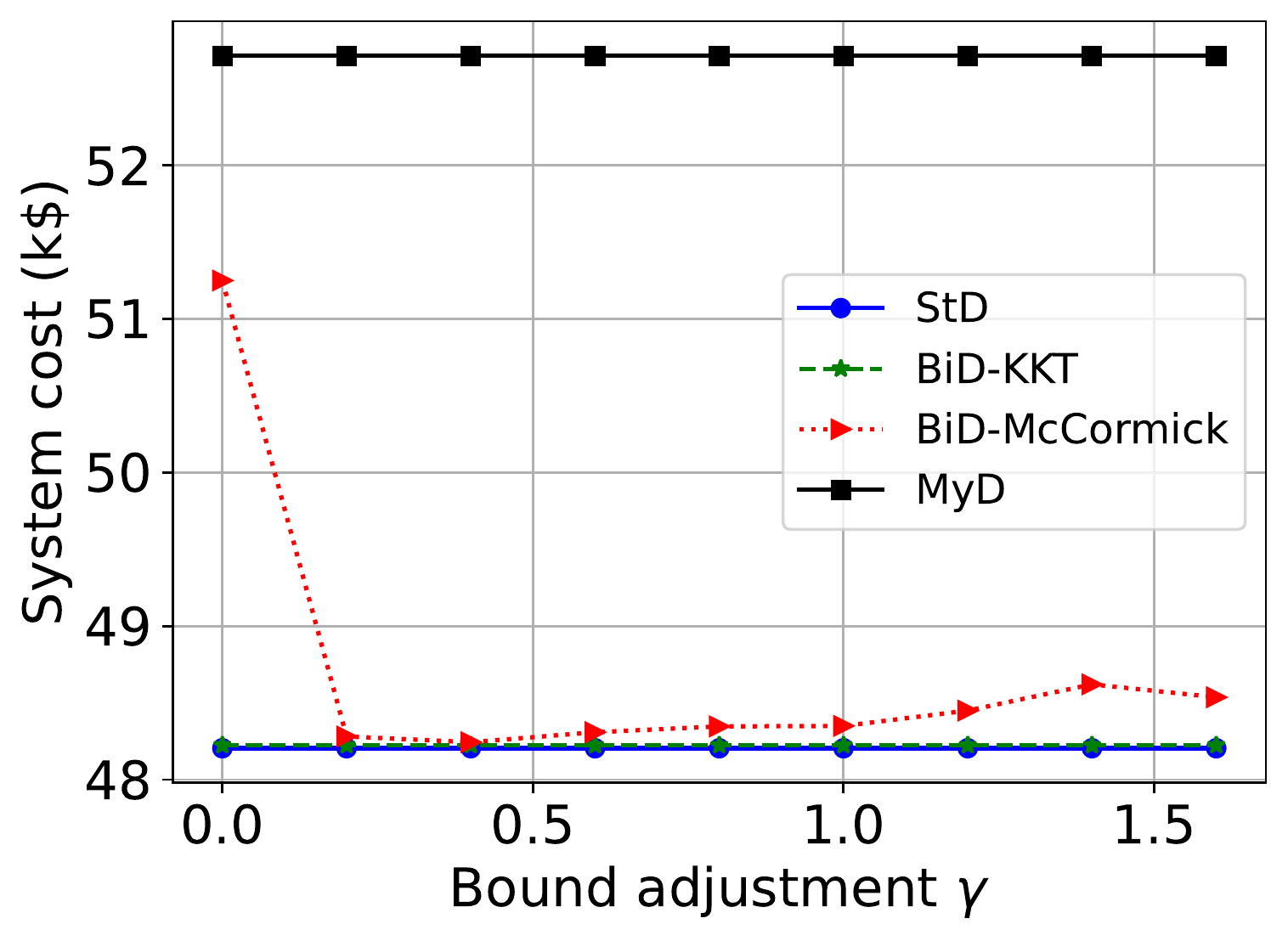}}}
	\hspace{-2ex}
	\subfigure[]{
		\raisebox{-2mm}{\includegraphics[width=1.67in]{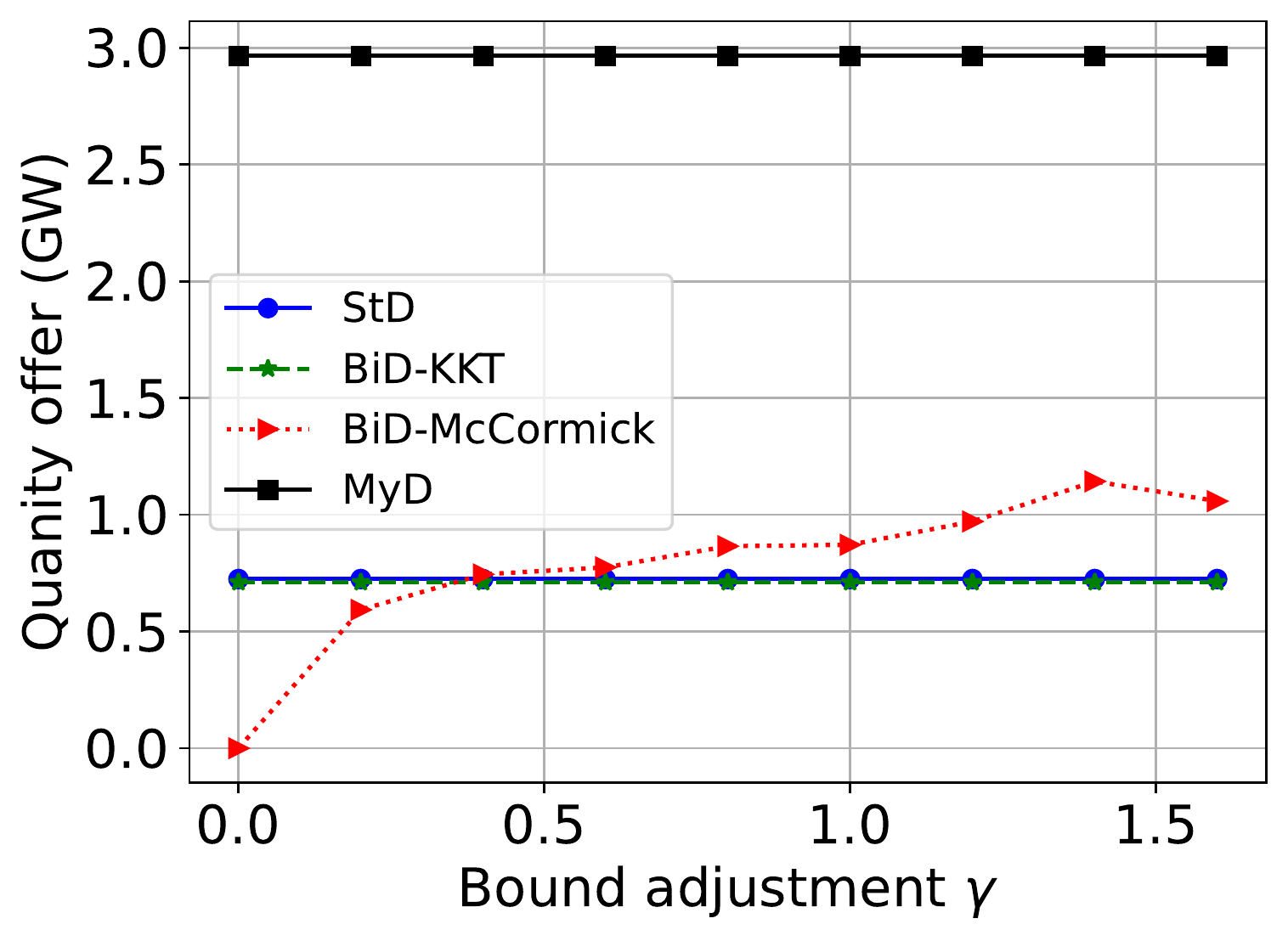}}}
	\vspace{-5mm}
	\caption{IEEE 118-bus system: (a) \small System cost; (b) Day-ahead aggregate VRES quantity. Both are functions of  bound adjustment $\gamma$.}
	\label{fig:bounds}
 \vspace{-2ex}
\end{figure}

Figures \ref{fig:bounds}(a) and (b) show system costs and DA VRES quantities, respectively, when we adjust the  upper bound parameter $\gamma$ in \eqref{eq:gamma} under \textit{BiD-McCormick}. Different curves represent \textit{StD}, \textit{BiD-KKT}, \textit{BiD-McCormick}, and \textit{MyD}, respectively. Note that \textit{BiD-KKT} refers to solving the bilevel problem directly using  KKT-condition-based MILP. To simulate a case of high VRES integration, we increase the capacity of wind farms so that the VRES penetration level reaches 70\%, and double the transmission capacity to accommodate this VRES expansion.

As we increase $\gamma$ in a wide range from 0.2 to 1.6, in Figure \ref{fig:bounds}(a), the system cost under \textit{BiD-McCormick} (red curve) is always very close to \textit{StD} (blue curve)  within the gap $0.7\%$, but lower than \textit{MyD} (black curve) by about 8\%. This shows that \textit{BiD-McCormick} can achieve very close performance to the least-cost benchmark \textit{StD}, which is also robust under the bound choice of  $\gamma$. As shown in Figure \ref{fig:bounds}(b), the DA scheduled VRES quantities under \textit{StD}, \textit{BiD-KKT}, and \textit{BiD-McCormick} are much more conservative than \textit{MyD} so as to avoid the shortage redispatch cost in the RT market.

In this small system, \textit{BiD-KKT} and \textit{-McCormick} can both compute solutions in about one minute.

\begin{figure}[t]
	\centering
	\includegraphics[width=2in]{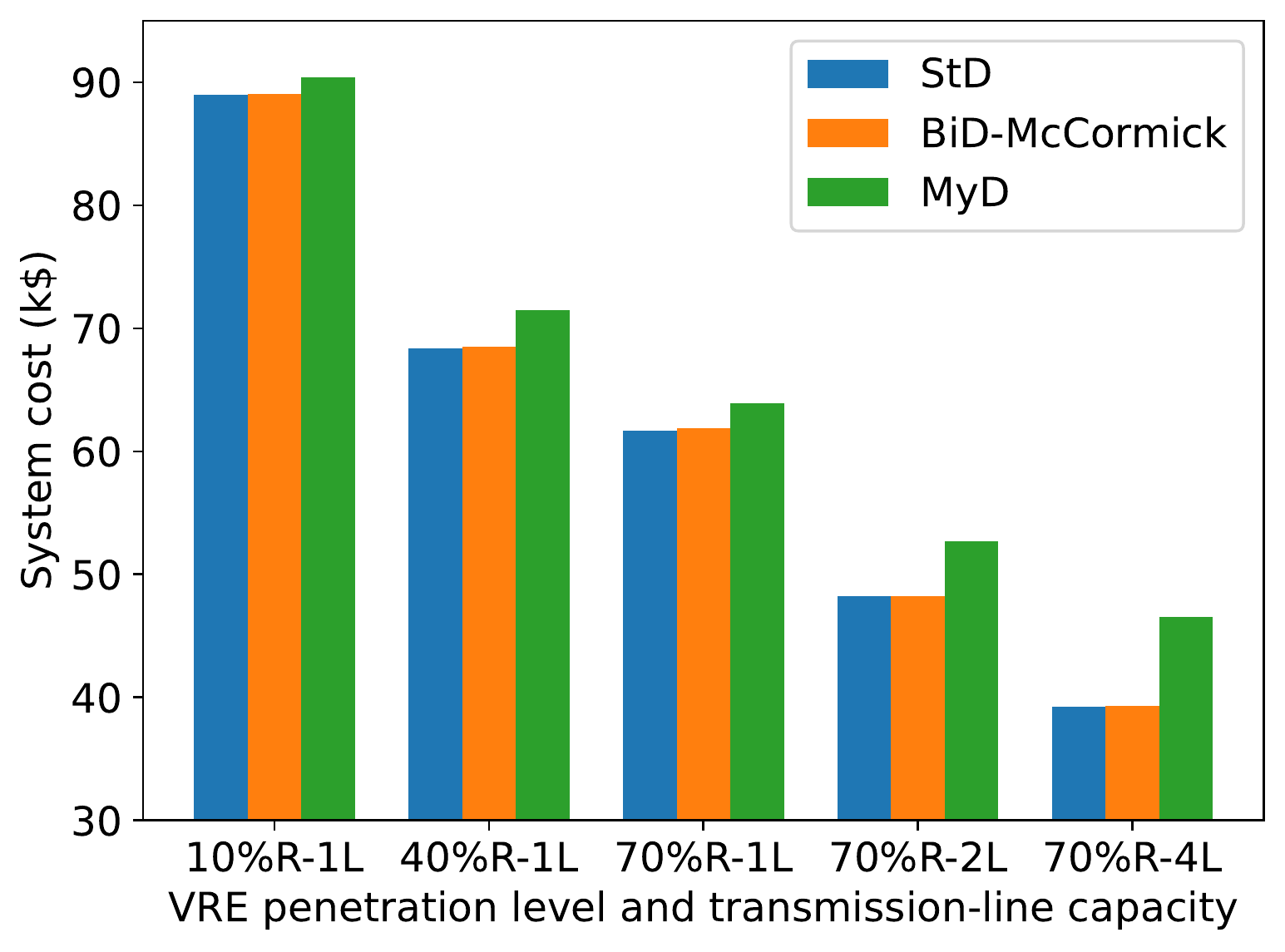}
	\vspace{-3ex}
	\caption{IEEE 118-bus system: System costs under different VRES penetration levels and transmission-line capacities.}
	\label{fig:comp}
 \vspace{-3ex}
\end{figure}

\subsubsection{System-cost gap between MyD and StD}

Figure \ref{fig:comp} shows the system-cost comparison between  \textit{StD},  \textit{BiD-McCormick}, and  \textit{MyD} under different settings of  $\mu \text{R}-\nu \text{L}$, where $\mu$  denotes the VRES penetration level and $\nu$ denotes the ratio of transmission-line capacity to the original one.

As shown in Figure \ref{fig:comp},  higher penetration levels of VRES and transmission capacity will reduce the system cost but increase the system-cost gap between \textit{MyD} and \textit{BiD-McCormick} or \textit{StD}. The intuition is that although higher integration of VRES will reduce the scheduling of high-cost conventional generations, it also brings more uncertainty to the system. In this case, our bilevel model reduces the system cost by over 15\% compared with  \textit{MyD}. Note that the performance is affected by various system parameters and we will provide more sensitivity analysis in future work.

\vspace{-0.5ex}
\subsection{Case study of the NYISO system}

We show that \textit{BiD-McCormick} can still compute accurate solutions in minutes on the large-scale NYISO system but \textit{BiD-KKT} cannot.

The tested  NYISO system has 1814 buses, 2264 transmission lines, 1564 loads, 345 conventional generation, and 14 wind farms.  The original number and capacity of wind farms are small, and only meet 4.4\% of the total demand. We generate forecast scenarios of wind energy following the truncated normal distribution within the capacity. To simulate a high VRES level, we increase the capacity of wind farms so that the VRES penetration level reaches 70\%, and upgrade the transmission capacity to 5 times, accordingly.

\begin{table}[bt!]
\caption{NYISO: System cost (\$1000) and computation time (second) for a varying number of uncertainty scenarios.}
\vspace{-3ex}
\label{tab:cpu_nyiso}
\begin{center}
\begin{tabular}{lcccccc}
\toprule
\multirow{2}{*}{\# of scenarios} & \multicolumn{2}{c}{10} & \multicolumn{2}{c}{20} & \multicolumn{2}{c}{50} \\
\cmidrule(lr){2-3}
\cmidrule(lr){4-5}
\cmidrule(lr){6-7}
 & cost & time & cost & time & cost & time \\
\midrule
\textit{MyD}     & 392.9 & 14 & 393.0 &  13 & 398.7 &  14  \\
\textit{BiD-McCormick} & 383.4 & 165 & 383.2 & 224 & 387.7 & 619  \\
\textit{BiD-KKT} & --- & $\gg2$h & --- & $\gg2$h & --- & $\gg2$h   \\
\textit{StD}   & 383.4 & 44 & 383.1 &  70 & 387.5 &  184 \\
\bottomrule
\end{tabular}
\end{center}
\vspace{-4ex}
\end{table}

Table \ref{tab:cpu_nyiso} presents the results of the system cost and computation time under \textit{MyD}, \textit{BiD-McCormick}, \textit{BiD-KKT},  and \textit{StD}. 
\textit{BiD-McCormick} can compute solutions within minutes. The computation time increases from 3 minutes to 10 minutes as the number of scenarios increases from  10 to 50. However, the conventional KKT method fails to give solutions within 2 hours. The system cost under \textit{BiD-McCormick} is almost the same as the ideal \textit{StD}. However, we also notice that the system cost reduction under \textit{StD} compared with \textit{MyD} is around 2.5\%, which is relatively modest under the high VRES penetration level. In future work, including ramping constraints and unit commitment decisions may increase this gap.

\vspace{-0.2ex}
\section{Conclusion}\label{section:conclusion}
In this work, we propose a computationally efficient mechanism to adjust the VRES quantity in the day-ahead market, which accounts for the uncertain real-time imbalance costs. The proposed scheme remains compatible with the existing sequential market-clearing structure. To facilitate computation, we propose a linear relaxation for the bilevel problem based on strong duality and McCormick envelopes. We conduct numerical studies on both IEEE 118-bus and NYISO systems. The results show that the proposed technique achieves good performance in terms of both accuracy and scalability on large-scale systems. Furthermore, the economic benefit of this bilevel VRES-quantity adjustment is more significant under higher VRES levels. The proposed model can serve as a decision-support tool for the market operator, which can provide a benchmark enabling market efficiency improvements. In future work, we will include ramping constraints and unit commitment decisions, and further improve the tightness of the McCormick envelope.

\bibliographystyle{ACM-Reference-Format}
\bibliography{storage}


\begin{thebibliography}{14}


\ifx \showCODEN    \undefined \def \showCODEN     #1{\unskip}     \fi
\ifx \showDOI      \undefined \def \showDOI       #1{#1}\fi
\ifx \showISBNx    \undefined \def \showISBNx     #1{\unskip}     \fi
\ifx \showISBNxiii \undefined \def \showISBNxiii  #1{\unskip}     \fi
\ifx \showISSN     \undefined \def \showISSN      #1{\unskip}     \fi
\ifx \showLCCN     \undefined \def \showLCCN      #1{\unskip}     \fi
\ifx \shownote     \undefined \def \shownote      #1{#1}          \fi
\ifx \showarticletitle \undefined \def \showarticletitle #1{#1}   \fi
\ifx \showURL      \undefined \def \showURL       {\relax}        \fi
\providecommand\bibfield[2]{#2}
\providecommand\bibinfo[2]{#2}
\providecommand\natexlab[1]{#1}
\providecommand\showeprint[2][]{arXiv:#2}

\bibitem[\protect\citeauthoryear{Boyd, Boyd, and Vandenberghe}{Boyd
  et~al\mbox{.}}{2004}]%
        {boyd2004convex}
\bibfield{author}{\bibinfo{person}{Stephen Boyd}, \bibinfo{person}{Stephen~P
  Boyd}, {and} \bibinfo{person}{Lieven Vandenberghe}.}
  \bibinfo{year}{2004}\natexlab{}.
\newblock \bibinfo{booktitle}{\emph{Convex optimization}}.
\newblock \bibinfo{publisher}{Cambridge university press}.
\newblock


\bibitem[\protect\citeauthoryear{Delikaraoglou and Pinson}{Delikaraoglou and
  Pinson}{2019}]%
        {delikaraoglou2019optimal}
\bibfield{author}{\bibinfo{person}{Stefanos Delikaraoglou} {and}
  \bibinfo{person}{Pierre Pinson}.} \bibinfo{year}{2019}\natexlab{}.
\newblock \showarticletitle{Optimal allocation of {HVDC} interconnections for
  exchange of energy and reserve capacity services}.
\newblock \bibinfo{journal}{\emph{Energy Systems}} \bibinfo{volume}{10},
  \bibinfo{number}{3} (\bibinfo{year}{2019}), \bibinfo{pages}{635--675}.
\newblock


\bibitem[\protect\citeauthoryear{Dvorkin, Delikaraoglou, and Morales}{Dvorkin
  et~al\mbox{.}}{2018}]%
        {dvorkin2018setting}
\bibfield{author}{\bibinfo{person}{Vladimir Dvorkin}, \bibinfo{person}{Stefanos
  Delikaraoglou}, {and} \bibinfo{person}{Juan~M Morales}.}
  \bibinfo{year}{2018}\natexlab{}.
\newblock \showarticletitle{Setting reserve requirements to approximate the
  efficiency of the stochastic dispatch}.
\newblock \bibinfo{journal}{\emph{IEEE Trans. on Power Systems}}
  \bibinfo{volume}{34}, \bibinfo{number}{2} (\bibinfo{year}{2018}),
  \bibinfo{pages}{1524--1536}.
\newblock


\bibitem[\protect\citeauthoryear{{Exizidis}, {Kazempour}, {Papakonstantinou},
  {Pinson}, {De Grève}, and {Vallée}}{{Exizidis} et~al\mbox{.}}{2019}]%
        {Exizidis2019market}
\bibfield{author}{\bibinfo{person}{L. {Exizidis}}, \bibinfo{person}{J.
  {Kazempour}}, \bibinfo{person}{A. {Papakonstantinou}}, \bibinfo{person}{P.
  {Pinson}}, \bibinfo{person}{Z. {De Grève}}, {and} \bibinfo{person}{F.
  {Vallée}}.} \bibinfo{year}{2019}\natexlab{}.
\newblock \showarticletitle{Incentive-Compatibility in a Two-Stage Stochastic
  Electricity Market With High Wind Power Penetration}.
\newblock \bibinfo{journal}{\emph{IEEE Transactions on Power Systems}}
  \bibinfo{volume}{34}, \bibinfo{number}{4} (\bibinfo{date}{July}
  \bibinfo{year}{2019}), \bibinfo{pages}{2846--2858}.
\newblock
\showISSN{0885-8950}
\urldef\tempurl%
\url{https://doi.org/10.1109/TPWRS.2019.2901249}
\showDOI{\tempurl}


\bibitem[\protect\citeauthoryear{for~a Smarter Electric Grid~(ICSEG)}{for~a
  Smarter Electric Grid~(ICSEG)}{[n.d.]}]%
        {ieee118}
\bibfield{author}{\bibinfo{person}{Illinois~Center for~a Smarter Electric
  Grid~(ICSEG)}.} \bibinfo{year}{[n.d.]}\natexlab{}.
\newblock \bibinfo{booktitle}{\emph{118 Bus Power Flow Test Case}}.
\newblock
\urldef\tempurl%
\url{https://icseg.iti.illinois.edu/ieee-118-bus-system/}
\showURL{%
Retrieved May 2, 2023 from \tempurl}


\bibitem[\protect\citeauthoryear{Greene}{Greene}{2022}]%
        {greene2022}
\bibfield{author}{\bibinfo{person}{Scott Greene}.}
  \bibinfo{year}{2022}\natexlab{}.
\newblock \bibinfo{booktitle}{\emph{{NYISO} network 2019}}.
\newblock \bibinfo{type}{{T}echnical {R}eport}.
  \bibinfo{institution}{University of Wisconsin-Madison}.
\newblock


\bibitem[\protect\citeauthoryear{{Kazempour}, {Pinson}, and
  {Hobbs}}{{Kazempour} et~al\mbox{.}}{2018}]%
        {Kazempour2018Market}
\bibfield{author}{\bibinfo{person}{J. {Kazempour}}, \bibinfo{person}{P.
  {Pinson}}, {and} \bibinfo{person}{B.~F. {Hobbs}}.}
  \bibinfo{year}{2018}\natexlab{}.
\newblock \showarticletitle{A Stochastic Market Design With Revenue Adequacy
  and Cost Recovery by Scenario: Benefits and Costs}.
\newblock \bibinfo{journal}{\emph{IEEE Transactions on Power Systems}}
  \bibinfo{volume}{33}, \bibinfo{number}{4} (\bibinfo{year}{2018}),
  \bibinfo{pages}{3531--3545}.
\newblock
\urldef\tempurl%
\url{https://doi.org/10.1109/TPWRS.2018.2789683}
\showDOI{\tempurl}


\bibitem[\protect\citeauthoryear{Kirschen and Strbac}{Kirschen and
  Strbac}{2018}]%
        {fundamentals}
\bibfield{author}{\bibinfo{person}{Daniel~S Kirschen} {and}
  \bibinfo{person}{Goran Strbac}.} \bibinfo{year}{2018}\natexlab{}.
\newblock \bibinfo{booktitle}{\emph{Fundamentals of power system economics}}.
\newblock \bibinfo{publisher}{John Wiley \& Sons}.
\newblock


\bibitem[\protect\citeauthoryear{McCormick}{McCormick}{1976}]%
        {mccormick1976computability}
\bibfield{author}{\bibinfo{person}{Garth~P McCormick}.}
  \bibinfo{year}{1976}\natexlab{}.
\newblock \showarticletitle{Computability of global solutions to factorable
  nonconvex programs: Part {I}—Convex underestimating problems}.
\newblock \bibinfo{journal}{\emph{Mathematical programming}}
  \bibinfo{volume}{10}, \bibinfo{number}{1} (\bibinfo{year}{1976}),
  \bibinfo{pages}{147--175}.
\newblock


\bibitem[\protect\citeauthoryear{Morales, Conejo, Liu, and Zhong}{Morales
  et~al\mbox{.}}{2012}]%
        {morales2012pricing}
\bibfield{author}{\bibinfo{person}{Juan~M Morales}, \bibinfo{person}{Antonio~J
  Conejo}, \bibinfo{person}{Kai Liu}, {and} \bibinfo{person}{Jin Zhong}.}
  \bibinfo{year}{2012}\natexlab{}.
\newblock \showarticletitle{Pricing electricity in pools with wind producers}.
\newblock \bibinfo{journal}{\emph{IEEE Transactions on Power Systems}}
  \bibinfo{volume}{27}, \bibinfo{number}{3} (\bibinfo{year}{2012}),
  \bibinfo{pages}{1366--1376}.
\newblock


\bibitem[\protect\citeauthoryear{Morales, Zugno, Pineda, and Pinson}{Morales
  et~al\mbox{.}}{2014}]%
        {morales2014electricity}
\bibfield{author}{\bibinfo{person}{Juan~M Morales}, \bibinfo{person}{Marco
  Zugno}, \bibinfo{person}{Salvador Pineda}, {and} \bibinfo{person}{Pierre
  Pinson}.} \bibinfo{year}{2014}\natexlab{}.
\newblock \showarticletitle{Electricity market clearing with improved
  scheduling of stochastic production}.
\newblock \bibinfo{journal}{\emph{European Journal of Operational Research}}
  \bibinfo{volume}{235}, \bibinfo{number}{3} (\bibinfo{year}{2014}),
  \bibinfo{pages}{765--774}.
\newblock


\bibitem[\protect\citeauthoryear{NYISO}{NYISO}{2021}]%
        {nyisowindrule}
\bibfield{author}{\bibinfo{person}{NYISO}.} \bibinfo{year}{2021}\natexlab{}.
\newblock \showarticletitle{Wind and Solar Resource Bidding, Scheduling,
  Dispatch, and Settlement}.
\newblock \bibinfo{journal}{\emph{NYISO: Technical Bulletin 154}}
  (\bibinfo{year}{2021}).
\newblock


\bibitem[\protect\citeauthoryear{REN21}{REN21}{2021}]%
        {renreport2021}
\bibfield{author}{\bibinfo{person}{REN21}.} \bibinfo{year}{2021}\natexlab{}.
\newblock \showarticletitle{Renewables 2021 Global Status Report}.
\newblock  (\bibinfo{year}{2021}).
\newblock


\bibitem[\protect\citeauthoryear{Viafora, Delikaraoglou, Pinson, Hug, and
  Holbll}{Viafora et~al\mbox{.}}{2020}]%
        {viafora2020dynamic}
\bibfield{author}{\bibinfo{person}{Nicola Viafora}, \bibinfo{person}{Stefanos
  Delikaraoglou}, \bibinfo{person}{Pierre Pinson}, \bibinfo{person}{Gabriela
  Hug}, {and} \bibinfo{person}{Joachim Holbll}.}
  \bibinfo{year}{2020}\natexlab{}.
\newblock \showarticletitle{Dynamic Reserve and Transmission Capacity
  Allocation in Wind-Dominated Power Systems}.
\newblock \bibinfo{journal}{\emph{IEEE Transactions on Power Systems}}
  (\bibinfo{year}{2020}).
\newblock


\end{thebibliography}

\begin{acks}
We would like to thank anonymous reviewers for their constructive
comments. 
This work has been supported  by ARPA-E Award No. DE-AR0001277.
\end{acks}

\appendix
\section{Notation}
\begin{table}[hbpt]
\allowdisplaybreaks
	\footnotesize
	\centering
	\caption*{Notation}
	\label{Tab:Nomec}
	\begin{tabular}{@{}lll@{}}
 \allowdisplaybreaks
		$(n,m)\in \Lambda$ & set &  set of transmission lines       \\
		$\omega \in \Omega$ & set & set of VRES production scenarios       \\
		$i \in \mathcal{I}$ & set & set of conventional generation units       \\
		$k \in \mathcal{K}$ & set & set of VRES power units       \\
		$n\in \mathcal{N}$ & set & set of buses       \\
		$\{\}_{n}$ & set & mapping of $\{\}$ into the set of buses      \\
		$\delta_{n}^{\text{DA}}$ & variable & day-ahead voltage angle at bus $n$   \\
		$\delta_{n\omega}^{\text{RT}}$ & variable & real-time voltage angle at bus $n$ in scenario $\omega$   \\
				$L_{n\omega}^{\text{sh}}$ & variable & shedding of load $n$ in scenario $\omega$    \\
		$P_{i}^{\text{C}}$ & variable & day-ahead schedule of conventional unit $i$    \\
		$P_{k}^{\text{W}}$ & variable & day-ahead schedule of VRES unit $k$     \\
		$P_{k\omega}^{\text{W,cr}}$ & variable & VRES power curtailment of unit $k$ in scenario $\omega$   \\
			$r_{i\omega}^{\text{U/D}}$ & variable & up-/downward redispatch of unit $i$ in scenario $\omega$    \\ 
      		${\lambda}_n^b$ & variable & dual variables associated with \eqref{eq:dabalance}  \\ 
             ${\lambda}_i^C$ & variable & dual variables associated with \eqref{eq:dac}  \\ 
            ${\lambda}_k^W$ & variable & dual variables associated with \eqref{eq:daw}  \\ 
            $\underline{\lambda}_{nm},\overline{\lambda}_{nm}$ & variable & dual variables associated with \eqref{eq:daline}  \\ 
   		$\bm{\lambda}$ & variable & all the  dual variables  for problem \eqref{eq:da}  \\ 
     $\tilde{\bm{\lambda}}$ & variable & all the  dual variables $\bm{\lambda}$  except those associated with \eqref{eq:dabalance}   \\ 
		$C_{i}$ & parameter & day-ahead energy price offer of unit $i$ \\ 
  	$C_{i}^{U/D}$ & parameter & real-time  up-/downward redispatch cost of unit $i$ \\ 
		$C^{\text{VoLL}}$ & parameter & cost of lost load  \\   
		$\overline{F}_{nm}$ & parameter & capacity of transmission line $(n,m)$  \\
	
		$L_{n}$ & parameter & demand of load located at bus $n$ \\   
		$\overline{P}_{i}$ & parameter & day-ahead quantity offer of unit $i$ \\ 
			$\widetilde{W}_{k\omega}$ & parameter & VRES power realization of unit $k$ in scenario $\omega$  \\
   		$\overline{W}_{k}$ & parameter & VRES power capacity of unit $k$  \\
     	$x_{nm}$ & parameter & reactance of transmission line $(n,m)$ 
	\end{tabular}
\end{table}

\section{Benchmarks}\label{sec:benchmark}

We consider two dispatch  benchmarks for \textit{BiD}. (i) \textit{Myopic dispatch (MyD):} Each VRES producer $k$ offers the bidding quantity at the expected value of the forecast, i.e.,   $W_k=\mathbb{E}_{\omega \in \Omega}[\widetilde{W}_{k\omega}]$. Then, the markets are cleared sequentially. (ii) \textit{Stochastic dispatch (StD):}
The system operator jointly schedules the day-ahead and real-time dispatches by minimizing the total expected costs in the two markets.

\setlength{\fboxsep}{6pt}
\setlength{\fboxrule}{1pt}
\fcolorbox{black}{gray!2}{%
   \parbox{0.85\columnwidth}{%
       \textbf{Problem \textit{StD}: Joint stochastic dispatch model}
\begin{subequations}
\begin{align*}
S^{\text{StD}}:=	{\text{min}} ~&
	f^{\text{DA}}(\Phi^{\text{DA}}) + \mathbb{E}_{\omega\in \Omega}\left[f_\omega^{\text{RT}}(\Phi_\omega^{\text{RT}})\right] \\
	\text{s.t.}	~&\Phi_\omega^{\text{RT}}\in \mathcal{X}_\omega^{\text{RT}}(\Phi^{\text{DA}}),~\forall \omega \in \Omega,\\
	&\Phi^{\text{DA}}\in \mathcal{X}^{\text{DA}}(\overline{\bm{W}}),\\
 \text{var:} ~&\Phi^{\text{DA}},~\Phi^{\text{RT}}.\notag
\end{align*} 
\vspace{-3.5ex}
\end{subequations}
}
}
\vspace{0.5ex}

The system costs always satisfy  $S^{\text{MyD}}\geq S^{\text{BiD}}\geq S^{\text{StD}}$.
The reason is that the solution of $\textit{MyD}$ is one feasible solution to Problem \textit{BiD} while the optimal solution to Problem \textit{BiD} is one feasible solution to Problem \textit{StD}. The two benchmarks serve as the upper bound and lower bound for the proposed bilevel model.

Notably,  while \textit{StD} attains the minimum dispatch cost, it does not follow economic dispatch in DA market clearing, making it hard to be directly implemented in the sequential market-clearing procedure. However, the proposed bilevel model can efficiently approximate the \textit{StD} solution while maintaining the sequential market structure.

\section{Bounds choice for McCormick envelope} \label{sec:bound}

It is important to choose proper lower and upper bounds for $ W_k $ and  $\overline{\lambda}_k^W$.  Note that we always have $ 0\leq W_k \leq \overline{W}_k$ and $0\leq \overline{\lambda}_k^W$ thanks to primal and dual feasibility conditions, respectively.  In the later numerical studies, for $W_k$, $\forall k\in \mathcal{K}$, we choose the bounds $\alpha_k^W=0$ and 
\begin{align}
  \beta_k^W=\gamma \cdot \mathbb{E}_{\omega}[\widetilde{W}_{k\omega}],\label{eq:gamma}
\end{align}
where we adjust $\gamma$ for the upper bounds around the mean value. For $\overline{\lambda}_k^W$, we let $\alpha_k^\lambda=0$ and $\beta_k^\lambda=\overline{\lambda}_k(\bm{0})$, where $\overline{\lambda}_k(\bm{0})$ denotes the dual solution of $\overline{\lambda}_k$ in the DA market when all VRES producers bid zero quantity. Since VRES have zero marginal cost in the DA market, the dual value $\overline{\lambda}_k(\bm{0})$ roughly indicates the upper bound for $\overline{\lambda}_k$. 

Although the selected lower and upper bounds for $ W_k $ and  $\overline{\lambda}_k^W$ are in a wide range, they can guarantee the system cost very close to \textit{StD} when $\gamma$ is within $[0.2,1.6]$ as shown in Section \ref{section:simulation}.

%
%
%
%
%
%

\end{document}